\documentstyle[a4,11pt,epsfig,twoside]{article}
\begin{document}
\setcounter{page}{1}
\title{Direct Searches of New Physics at CLIC}
\author{M. Battaglia\thanks{e-mail address: Marco.Battaglia@cern.ch}, 
        A. De Roeck\thanks{e-mail address: Albert.De.Roeck@cern.ch}
\\
\\
         {\it CERN, Geneva, Switzerland}
\\
\\
         T. Rizzo
        \thanks{e-mail address: rizzo@SLAC.Stanford.edu}
\\
\\
         {\it SLAC, 
Stanford University, Stanford, California 94309 USA}
}

\date{}
\maketitle
\begin{abstract}
The multi-TeV $e^+e^-$ collider CLIC may allow for the direct study of 
new neutral gauge bosons
or  Kaluza-Klein states in the TeV range. We discuss some of the 
experimental aspects for the study of such resonances. 
Further we  discuss briefly the effects of soft branes in scenarios 
with Large Extra Dimensions, 
and the production of Black Holes at CLIC.

\end{abstract}

\section{Introduction}

CLIC (Compact LInear Collider) is a concept for a linear $e^+e^-$ 
collider with centre of mass (CMS)
 energies in the range of 3 to 5 TeV and a luminosity
of 10$^{35}$ cm$^{-2}$s$^{-1}$. The accelerating principle is based
on two-beam acceleration~\cite{clic}, which is presently still in 
an experimental stage but has booked quite a few important successes
in the last few years~\cite{clicplenar}.
To reach such high luminosities CLIC operates in a high
beamstrahlung environment, which distorts the luminosity spectrum and 
leads to important backgrounds, mostly $\gamma\gamma$ collisions, 
in the interaction region. Most studies in this paper were performed using 
tools which take into account these effects, as well as a detector
response of a TESLA-like detector with slightly modified interaction and
mask region~\cite{marco}. 

Several 
  different realisations of New Physics lead to  new vector resonances 
  and other phenomena for which the potential of CLIC has been studied.
In this paper we discuss examples of  direct searches for New Physics at CLIC:
the observation of a new Z' like resonance,
KK tower production, large extra dimensions with soft branes and the 
production of black holes. Results on indirect searches are discussed in
~\cite{decurtis}. Susy topics at CLIC are discussed in~\cite{gruwe}.

\section{Direct Production of New Gauge Bosons}

\begin{figure}[tbh]
\epsfig{file=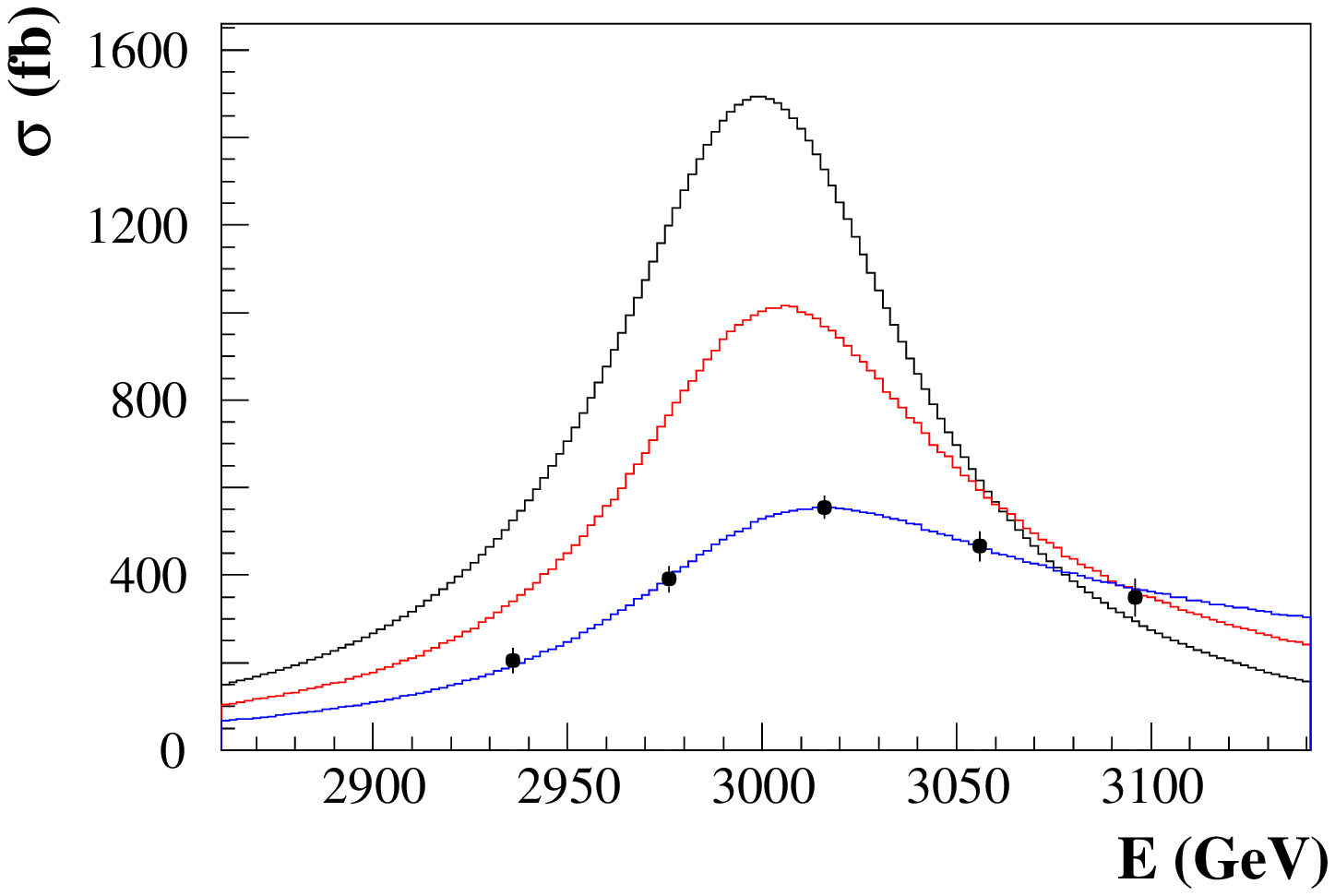,width=6cm,clip}
\epsfig{file=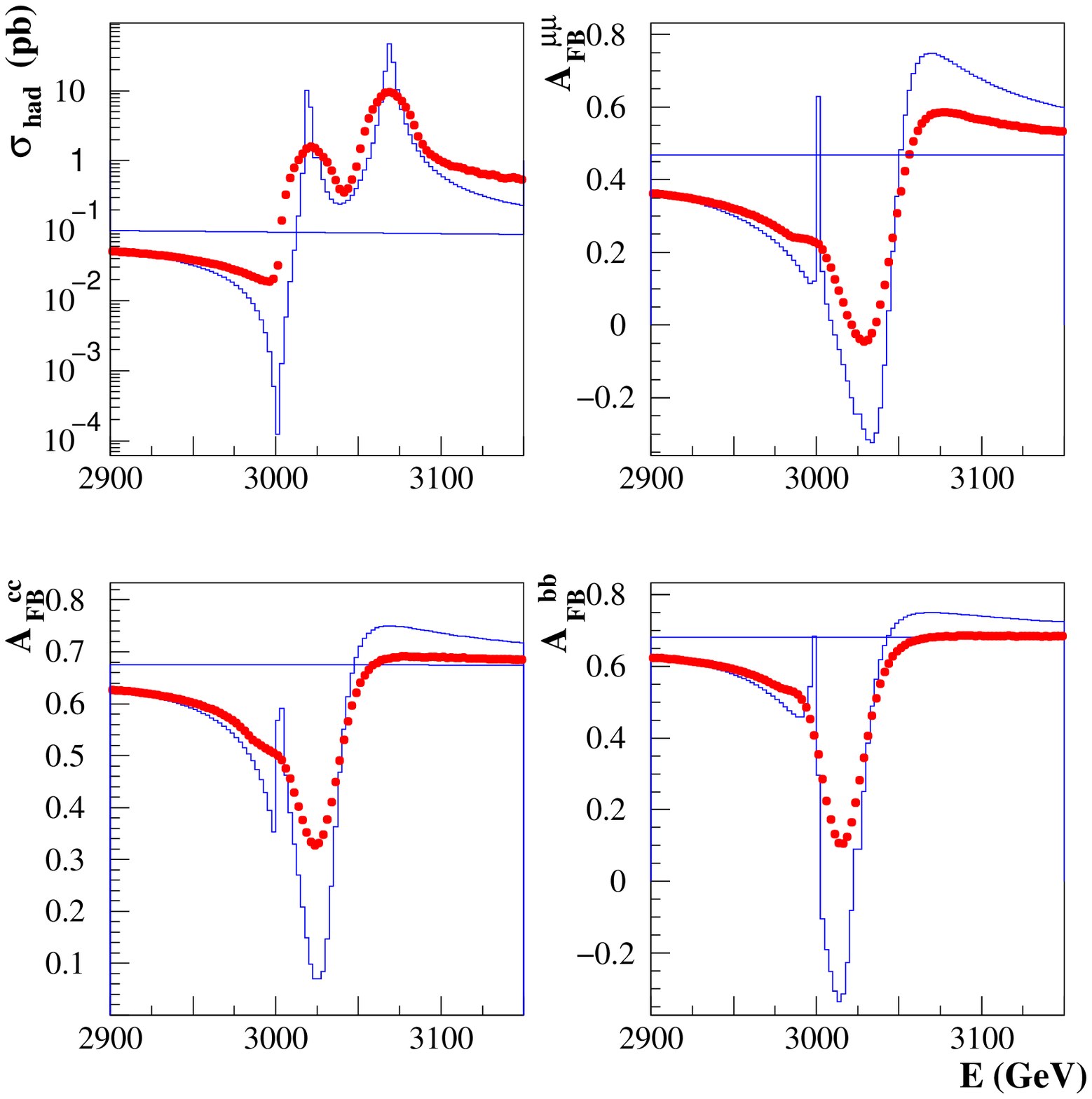,bbllx=60,bblly=140,bburx=550,bbury=640,width=6.0cm,clip}
\caption{(a) $Z'_{SSM} \to \ell^+ \ell^-$ resonance profile obtained by
energy scan. The Born production cross-section, the cross
section with ISR included and that accounting for the CLIC 
luminosity spectrum (CLIC.01) and tagging criteria are shown.
(b) Hadronic cross section (upper left)  and $\mu^+\mu^-$ (upper
right), $c \bar c$ (lower left) and $b \bar b$ (lower right)
forward-backward asymmetries at energies around 3~TeV. The
continuous lines represent the predictions for the D-BESS
model with $M$ = 3~TeV and $g/g''=0.15$, the flat lines the
SM expectation and the dots the observable D-BESS signal
after accounting for the CLIC.02 luminosity spectrum.}
\begin{picture}(0,0)
\put(140,230){\bf(a)}
\put(200,290){\bf(b)}
\end{picture}
\label{fig1}
\end{figure}

CLIC is an ideal collider to study 
the production of a new neutral gauge boson if its mass 
is within the kinematical reach. Similar to LEP, it will be able to 
make precision measurements of the boson properties. The results
of a scan over a Z' resonance~\cite{marco1}, assumed to have SM couplings, 
for one year of running and for two different assumptions for the
luminosity spectra are given in Table.~\ref{tab1}. 
The CLIC.01 is broader but delivers more integrated luminosity than
the CLIC.02 spectrum.
The pseudo data, shown in Fig.~\ref{fig1}, 
are shared over 5 data points
and $M_{Z'}$, $\Gamma_{Z'}/\Gamma_{Z^0}$
and $\sigma_{peak}$ have been extracted from a $\chi^2$ fit to the
predicted cross section behaviour for different mass and width values.
The relative statistical
accuracies are found to be better than 10$^{-4}$ on the mass and $5 \times
10^{-3}$ on the width. In the case of wide resonances, there is an advantage 
in employing the broader luminosity spectrum, CLIC.01. 
 Sources of systematics from the knowledge of
the shape of the luminosity spectrum have also been estimated. In
order to keep $\sigma_{syst} \le \sigma_{stat}$ it is necessary to
control $N_{\gamma}$ to better than 5\% and the fraction of
collisions at $\sqrt{s} < 0.995 \sqrt{s_{0}}$ to about 
1\%.

\begin{table}[t]
\begin{center}
\begin{tabular}{|l|c|c|c|}
\hline
Observable & Breit Wigner & CLIC.01 & CLIC.02 \\ \hline
$M_{Z'}$ (GeV) & 3000 $\pm$ .12  & $\pm$ .15 &  $\pm$ .21 \\
$\Gamma_{Z'}/\Gamma_{Z^0}$ & 1. $\pm$ .001 & $\pm$ .003  & $\pm$ .004 \\
$\sigma^{eff}_{peak}$ (fb) & 1493 $\pm$ 2.0 & 564 $\pm$ 1.7 & 669 $\pm$ 2.9 \\
\hline
\end{tabular}
\caption{Results of the fits for the cross section scan of a $Z'_{SSM}$ 
obtained 
by assuming no radiation and ISR with the effects of 
two different choices of the 
CLIC luminosity spectrum.}
\label{tab1}
\end{center}
\end{table}

Models based on strong electro-weak symmetry breaking often predict
several additional gauge bosons. Here we take as an example the degenerate 
BESS model~\cite{dbess} which introduces two new triples of gauge bosons
($L^{\pm},L_3)$ and $(R^{\pm}, R_3$) which are almost degenerate in mass.

The ability to identify the model distinctive 
features has been studied using the production cross section and the
 flavour dependent
forward-backward asymmetries, for different values of $g/g''$, where
 $g''$ is the new gauge coupling constant.
 The resulting 
distributions are shown in Fig.~\ref{fig1} for the case of the narrower 
CLIC.02 beam parameters. A characteristic feature of the cross 
section distributions is
the presence of a narrow dip, due to the interference 
of the $L_3$, $R_3$ resonances with the $\gamma$ and 
$Z^0$ and to cancellations of the 
$L_3$, $R_3$ contributions. 
Figure~\ref{fig1} shows that the effect is still visible 
after the smearing from  the
luminosity spectrum.  With realistic assumptions and 1~ab$^{-1}$ of data, 
CLIC  will be able to resolve the two narrow resonances 
for values of the coupling 
ratio $g/g''>$~0.08, corresponding to a mass splitting $\Delta M$ = 13~GeV 
for
$M=$ 3~TeV, and to determine $\Delta M$ with a statistical 
accuracy better than 
100~MeV.

\section{Extra Dimensions in the Randall-Sundrum Model}

For the past few years the phenomenology of 
scenarios with extra dimensions has been 
explored at the TeV scale. These theories aim to solve the hierarchy problem
by bringing the Gravity scale  closer to the  Electroweak scale.

In the extra-dimension scenario proposed by 
Randall-Sundrum(RS)~\cite{RS}  the hierarchy between the Planck 
and the Electroweak scale is generated by an exponential function called
'warp factor'. 
This model predicts Kaluza-Klein graviton resonances with both
weak scale masses and couplings to matter in the TeV range.

\begin{figure}[tbp]   
\begin{center} 
\epsfig{file=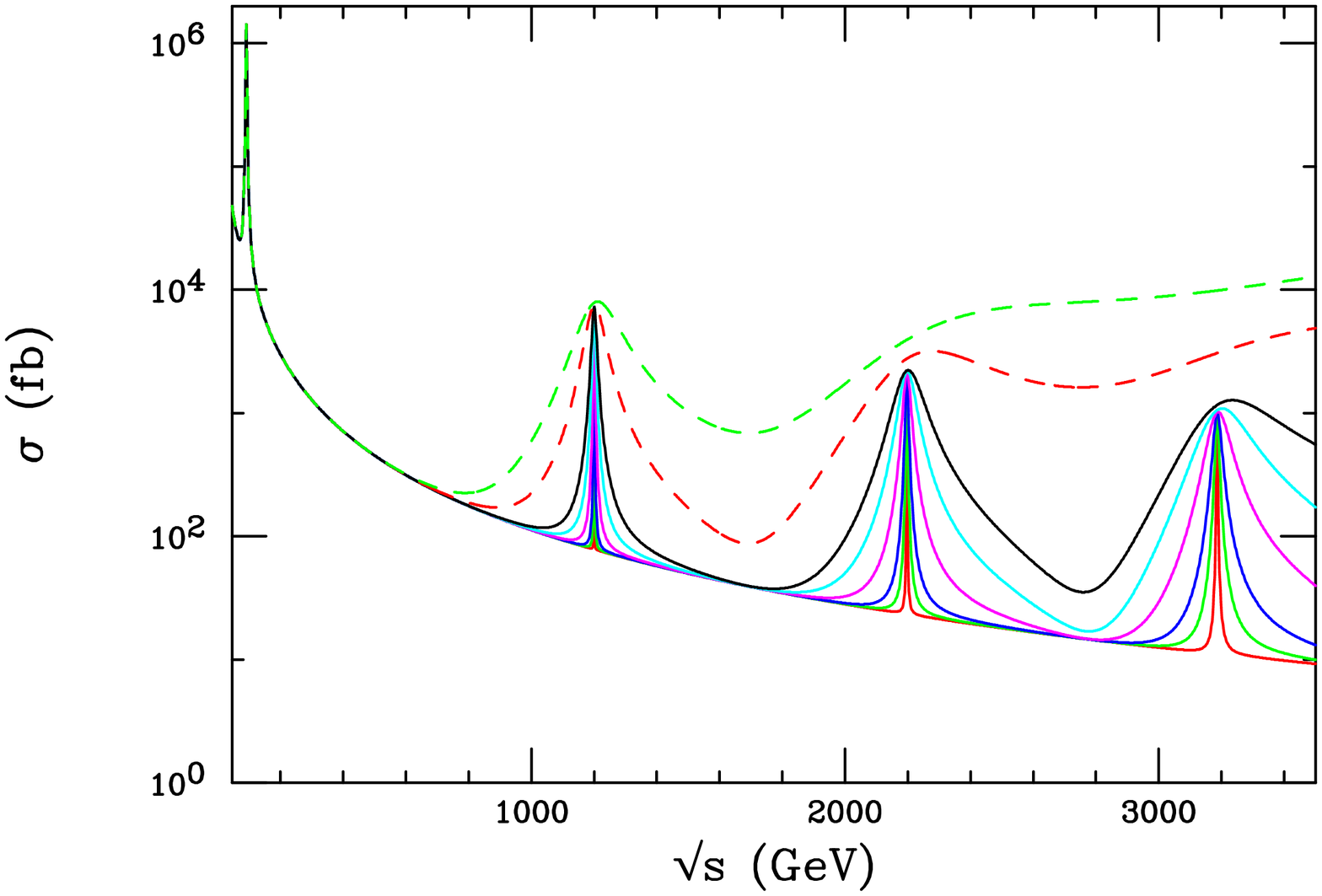,bbllx=90,bblly=90,bburx=550,bbury=720,width=5cm,angle=90,clip}
\epsfig{file=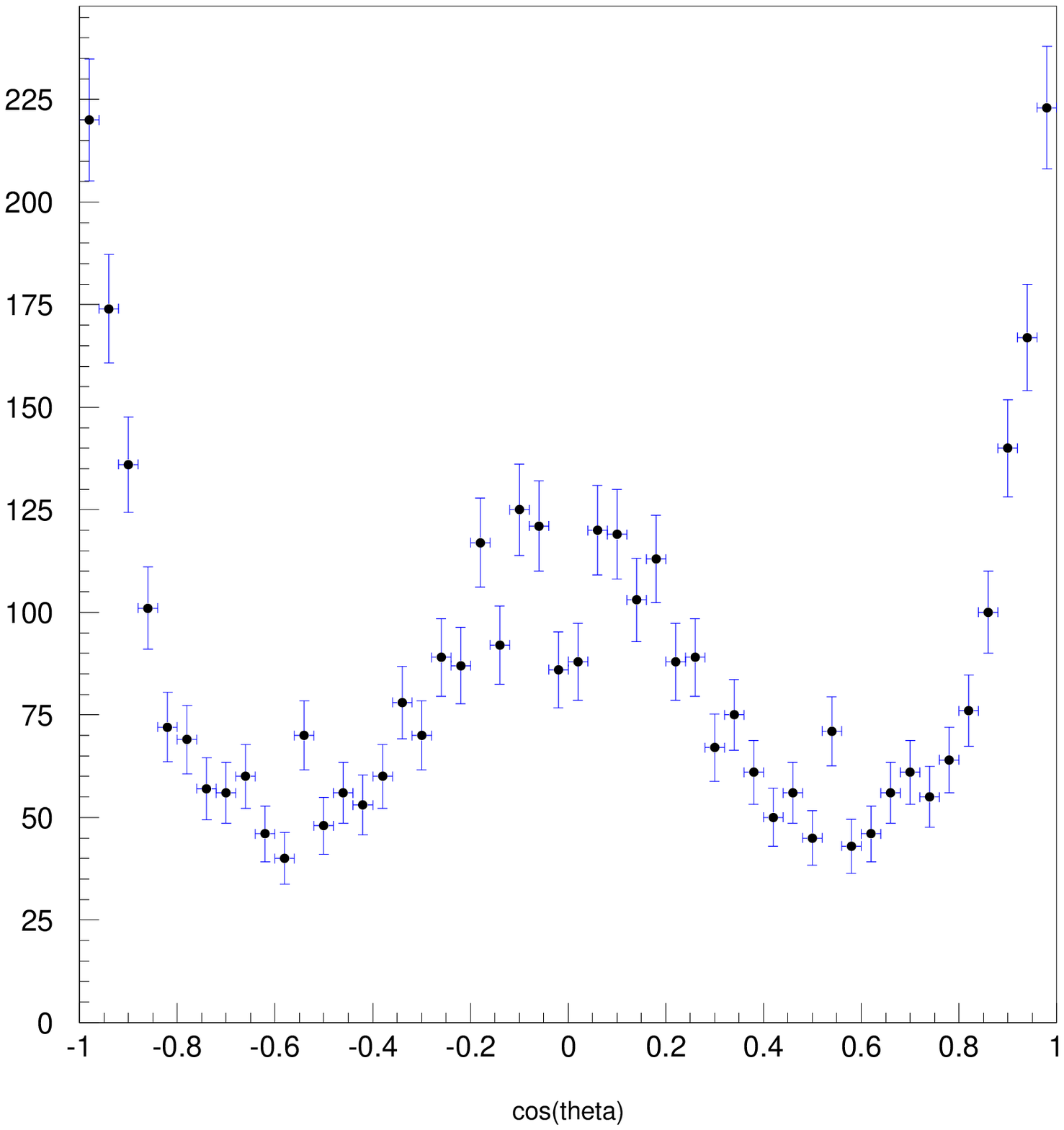,bbllx=0,bblly=0,bburx=565,bbury=565,width=5cm}
\caption{
(Left) KK graviton excitations in the RS model 
produced in the process $e^+e^-\to \mu^+\mu^-$. From the most narrow to widest 
resonances the curves are for $c$ in the range 0.01 to 0.2. 
(Right) Decay angle distribution of the muons from 
 $G_3 (3200 \rm{GeV})\rightarrow \mu\mu$.}
\label{fig2}
\end{center}
\end{figure}

An example of a spectrum for $e^+e^-\rightarrow \mu^+\mu^-$ is shown in 
Fig.~\ref{fig2}, for different values of 
the parameter $c$ which
controls the effective coupling
strength of the graviton and thus the width of the resonances.
 The cross sections are huge and the 
signal cannot be missed at a LC with sufficient CMS  energy.
The resonance spectrum was chosen such that the first resonance $G_1$
has a mass $M$ around 1200 GeV, just outside the reach of a
TeV class LC, and consequently the mass of the third resonance 
$G_3$ will be around
3200 GeV, as shown in Fig.~\ref{fig2}.
The CMS energy for the $e^+e^-$ collisions of CLIC was taken to be 3.2 TeV
in this study.
Mainly the muon and photon decay modes of the graviton have been studied.
The  events used  to reconstruct the $G_3$ resonance signal were selected
either via 
two muons or two $\gamma$'s with $E> 1200$ GeV and $|\cos\theta| <0.97$.
The background from overlaid two-photon events -- on average four events
per bunch crossing --is typically important only for angles below 120 mrad, 
i.e. outside the considered  signal search region.

First we study the precision with which we can measure the shape, i.e.
the  $c$ and  $M$ 
parameters, of  the observed  new resonance, similar to the Z',
for an integrated luminosity of  1 ab$^{-1}$.
The precision with which the cross sections are measured allows one to 
determine $c$ to  0.2\%, and $M$ to better than 0.1\%. 
Next we determine some key  properties of the new resonance:
  the spin and the ratio of decay modes.
The graviton is a spin-two object. Fig.~\ref{fig2} shows the 
decay angle of the  fermions $G\rightarrow \mu\mu$ 
for the $G_3$ graviton,  for 1  ab$^{-1}$ of data,
 including CLIC machine background.
The typical spin-two structure of the decay angle 
 of the resonance is clearly visible.
For gravitons as proposed in~\cite{dhr,rizzo3} one expects
{$BR(G \rightarrow \gamma\gamma)/BR(G \rightarrow \mu\mu)$ = 2.}
With the present detector  simulation we get
efficiencies in mass peak ($\pm 200$  GeV) 
of 84\% and 97\% for detecting the muon and photon decay modes, respectively.
With cross 
sections of $O(pb)$, $\sigma_{\gamma\gamma}$ and $\sigma_{\mu\mu}$
can be determined to better than a per cent.
Hence the ratio $BR(G\rightarrow \gamma\gamma)/BR(G\rightarrow
\mu\mu)$ can be determined to an accuracy of 1\%  or better.
Finally, if the CMS energy of the collider is large 
enough to produce the first three resonances states, one has 
the intriguing possibility to measure the graviton 
self-coupling via the $G_3\rightarrow G_1G_1 $ decay~\cite{rizzo3}.
The dominant decay mode will be
$G_1 \rightarrow$ gluon-gluon or $q\bar q \rightarrow$ two jets.
It has been shown~\cite{albert} that four jet events 
 can be used to reconstruct $G_1$ resonances in the $G_3$ decay
with no significant distortion of the background.

\begin{figure}[htp]   
\begin{center} 
\epsfig{file=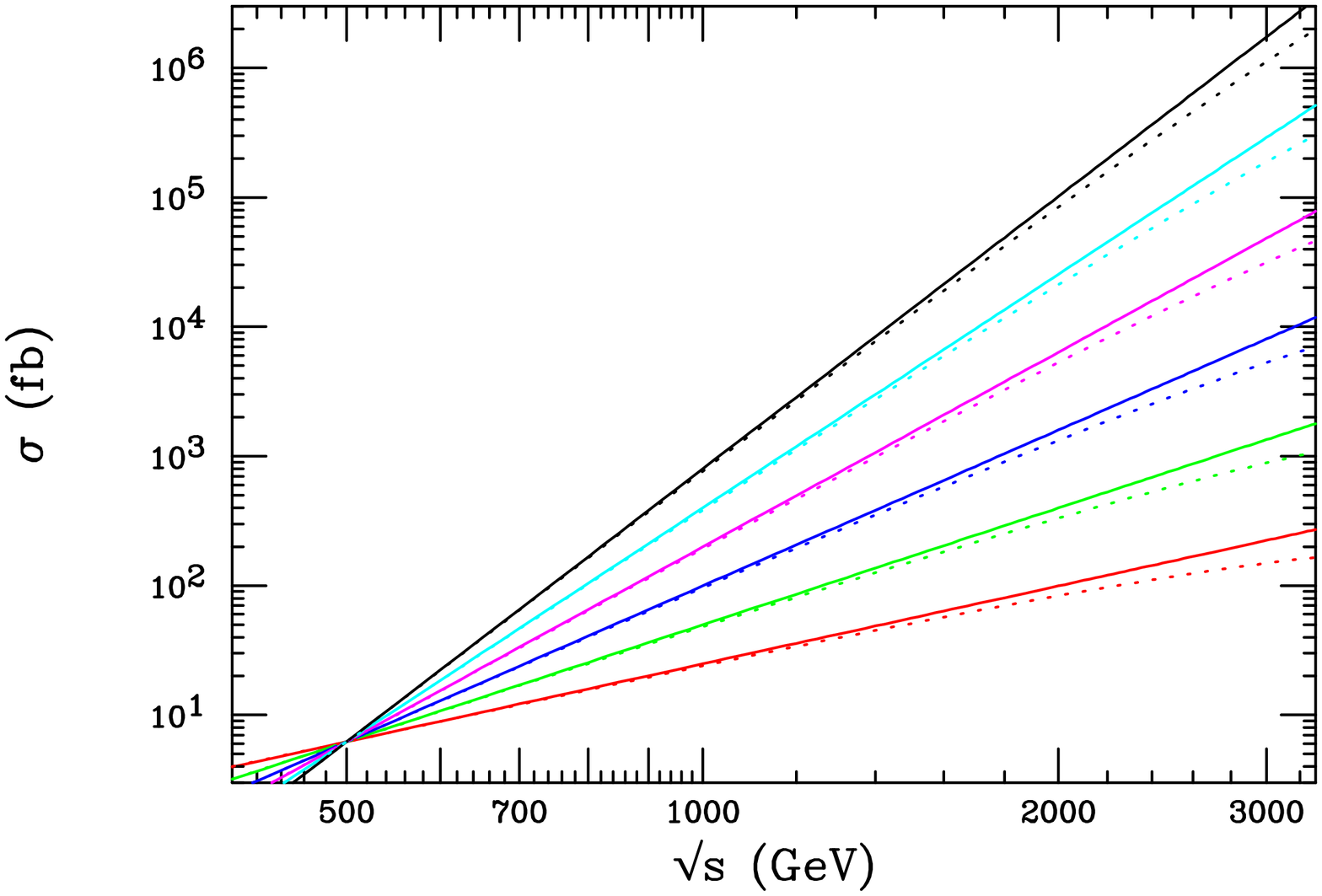,bbllx=0,bblly=100,bburx=540,bbury=700,width=5cm,angle=90}
\epsfig{file=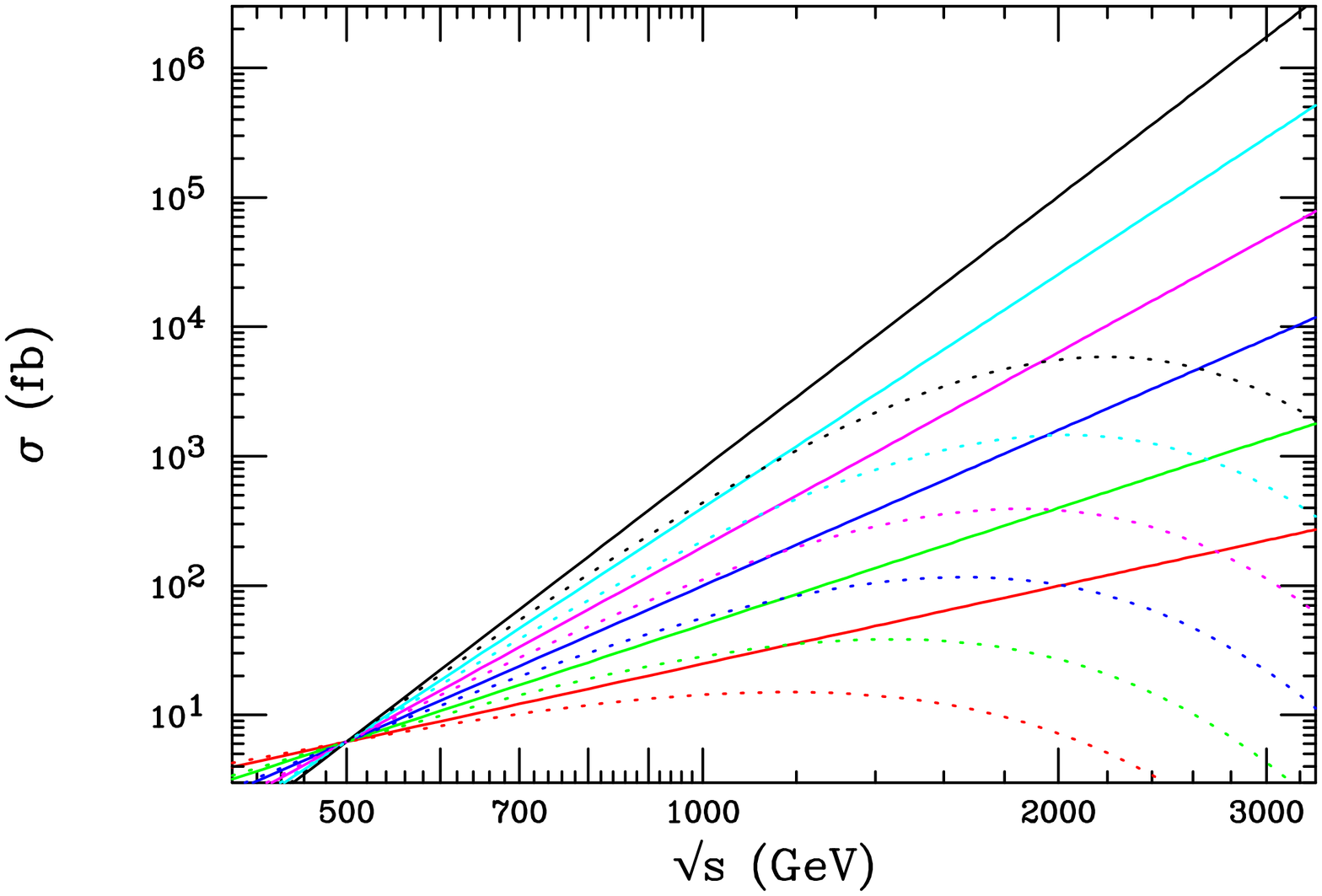,bbllx=0,bblly=100,bburx=540,bbury=700,width=5cm,angle=90}
\caption{The cross section for $e^+e^- \rightarrow G\gamma$ with cuts as
described in the text, for rigid (solid lines) and soft (dashed lines)
branes. The different curves correspond, from bottom to top, to different
number of extra dimensions $\delta$ = 2,3,4,5,6 and 7.
Left, for $\Delta = 4$ TeV, right for $\Delta= 1 $TeV.}
\label{fig3}
\end{center}
\end{figure}

\section{Large Extra Dimensions with Soft Branes}
Scenarios with large extra dimensions have been studied for 
TeV class linear colliders,
e.g. TESLA~\cite{teslaTDR} and are searched for typically in the channel
$e^+e^- \rightarrow \gamma +G$.
One of the acclaimed advantages of a LC w.r.t. the LHC is that by 
measuring the $\gamma G$ cross section at different center of mass energies
one can disentangle the Planck scale and the number of extra dimensions
$\delta$ simultaneously as is shown in Fig.~\ref{fig3} by the solid lines.
The cross sections are calculated for  cuts similar to corresponding to those 
in\cite{teslaTDR}:
$\sin \theta_{\gamma} > 0.1 $, $p^{\gamma}_t > 0.06E_{beam}$
and $x_{\gamma} < 0.65$.
The cross sections are normalized such that for each value of $\delta$,
the scale, $M_D$, is chosen to give the same cross section 
at 500 GeV. 

These predictions  however assume the branes are rigid. 
Allowing for flexible branes instead~\cite{wells} introduces a new dependence
on a  parameter $\Delta$, the softening scale, which is  related to the brane 
tension. The dashed lines in Fig.~\ref{fig3} shows the effect of $\Delta$ for 
values of 4 TeV  and 1 TeV respectively.

For a brane tension of 4 TeV the effect on the cross section is rather small.
A collider in the range of 0.5-1 TeV would not be sensitive to
the effect and thus $M_D$, and $\delta $ can be 
disentangled. However,  at CLIC the 
cross sections are 30-40\% lower  than expected, allowing
to observe the softness
of the brane.
For a brane with tension of 1 TeV the effect is more spectacular.
For the example given here a lower energy
 LC when  measuring cross sections only
at 0.5 TeV and 1 TeV  would get fooled and  
extract a wrong $\delta, M_D$. Extending the range in the 
multi-TeV region again will allow this effect to be observed in its full drama.
The cross section of the background channel
 $ee\rightarrow ee\gamma$ with the cuts as listed above
is 16 fb at 3 TeV, which sets the scale for the detectability of a signal:
for $\delta $ = 2 or 3 the signal event rate at 3 TeV gets too small
for such a soft brane scenario.

\section{Black Holes}
If the fundamental Planck scale is in the TeV range
then a possible consequence would
be  that
black holes (BH) could be produced in the multi-TeV range.
The cross section is  expected to be very large:
$\sigma = \pi R_s^2 \sim 1 $ TeV$^{-2}\sim O(100) $ pb,
where $R_S$ is the  Schwarzschild Radius.
If $\sqrt{s}_{e^+e^-} >M_{BH}>M_{\rm Planck}$ then the 
collider becomes a  black hole factory.
The lifetime of such a black hole is of order 
 $ \sim 10^{-25}-10^{-27}$ sec, and hence the black hole will evaporate before
it could possibly `attack' any detector material. 

\begin{figure}[htbp]
\begin{center}
\epsfig{file=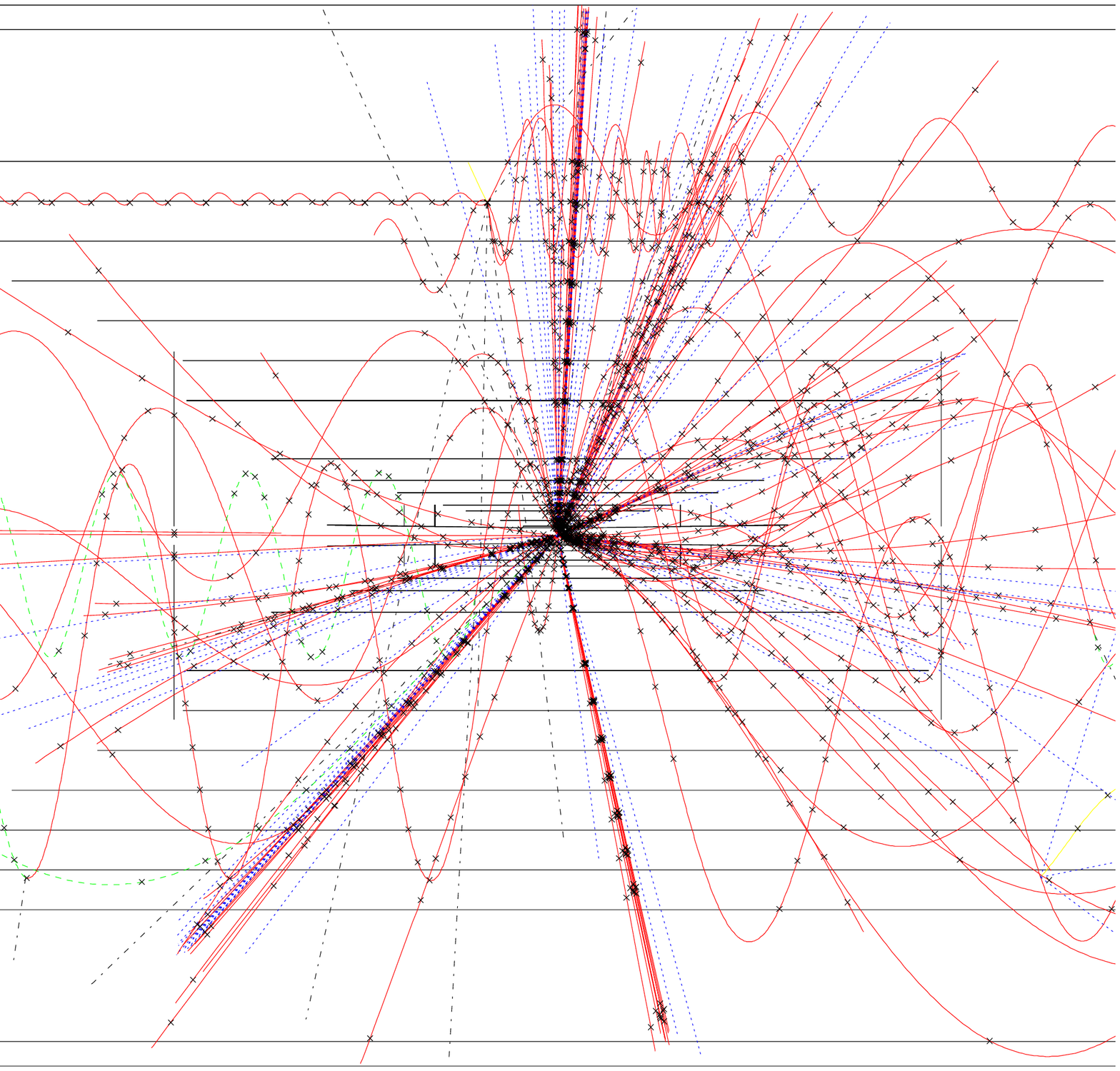,bbllx=0,bblly=0,bburx=590,bbury=570,width=6cm,clip=}
\epsfig{file=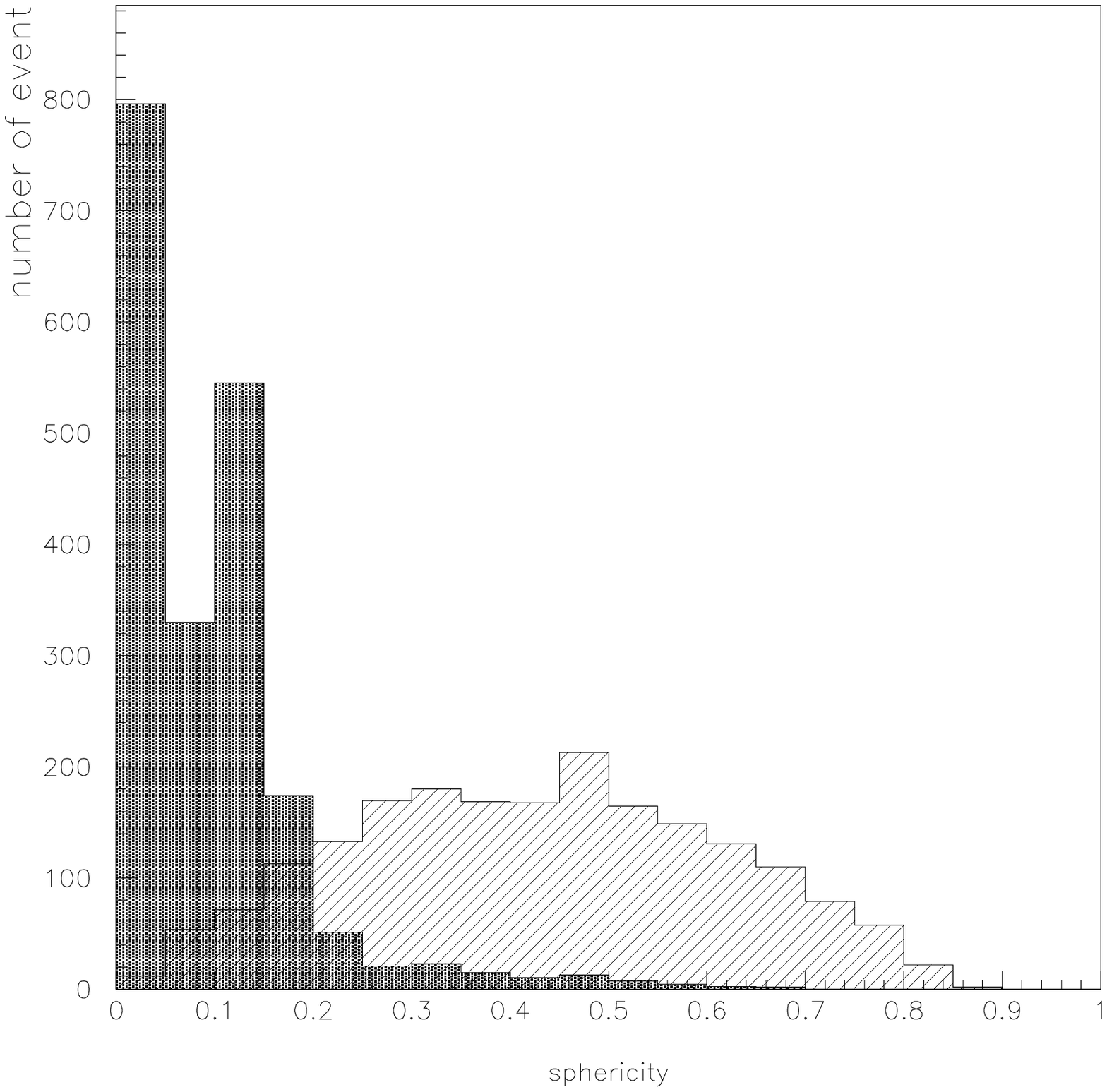,bbllx=0,bblly=130,bburx=540,bbury=700,width=6cm,clip=}
\end{center}
\caption{Black hole production in the CLIC detector.
(Left) example of an event, (Right) sphericity distribution
for 2 and 4 fermion events (full histogram) and black holes
(hatched histogram).}
\label{fig4}
\end{figure}

The decay of a black hole can be very complex and involve several
stages~\cite{thomas,landsberg}. If the dominating mode is 
 Hawking radiation then all 
particles (quarks, gluons, gauge bosons, leptons) are expected to 
be produced democratically, with e.g. a ratio 1/5 
between leptonic and hadronic activity. 
The multiplicity is expected to be large.
The production and decay process
have been included in the PYTHIA generator\cite{landsberg}.
Fig.~\ref{fig4}a shows a black hole event produced in a detector
at CLIC, leading to spectacular multi-jet and lepton/photon signals.
Black holes will be easily detected at CLIC due to their energetic
leptons and photons, and the more spherical event shape.
As an example Fig.~\ref{fig4}b shows, for 3 extra dimensions and 
a scale of 2 TeV, the sphericity of the events measured
after detector simulation and addition of 5 bunch crossings of $\gamma\gamma$
background for black holes and annihilation events. 

If this scenario is realized in Nature, black holes will be produced
at high rates at the LHC. CLIC would be very instrumental 
in providing 
precise measurements.  For example, it could be used to test Hawking 
radiation and 
extract the number of underlying extra dimensions.
The large production cross section, low backgrounds and little missing 
energy would make BH production and decay a perfect laboratory to study 
strings and quantum gravity in the lab.

\section{Conclusions}

The direct production of new gauge bosons and 
KK excitations in the TeV range was studied for CLIC.
The expected backgrounds and the smeared luminosity spectrum of CLIC 
do not 
 prevent to make precision measurements of the model parameters.
In particular for the RS model it was shown that the key discriminating 
properties of these resonances can be reconstructed and the underlying
model parameters can be determined precisely. 
Furthermore, CLIC will be very instrumental for the study of rigidness of 
branes in scenarios with large ED's,  and the study of 
black holes if the Planck scale is below the CMS energy of the machine.
Even though the LHC will likely detect most of such new phenomena, if these
exist, 
CLIC data will be needed to fully understand their nature and properties.

\begin{center}
{\it

We  would like to thank 
S. De Curtis, D. Dominici, J. Hewett and  G. Landsberg together with
whom part of this work was done.}

\end{center}




\end{document}